\documentclass[prb,twocolumn]{revtex4}

\pdfoutput=1

\usepackage[T1]{fontenc}
\usepackage{graphicx}
\usepackage{amsmath}
\usepackage{times}
\usepackage{amssymb,amsmath} % for math
\usepackage{epsfig}

\newcommand{\epsenv}{\varepsilon_{\rm env}}

\begin{document}

\title{Binding energies of exciton complexes
in transition metal dichalcogenide monolayers
and effect of dielectric environment}
\author{Ilkka Kyl\"anp\"a\"a$^1$ and Hannu-Pekka Komsa$^2$}

\affiliation{
$^1$Department of Physics, Tampere University of Technology,
P.O. Box 692, FI-33101 Tampere, Finland
}
\affiliation{
$^2$COMP, Department of Applied Physics, Aalto University,
P.O. Box 11100, 00076 Aalto, Finland
}

\date{\today}

\begin{abstract}
%Binding of exciton complexes and the effect of dielectric environment
%in transition metal dichalcogenides is studied by means of 
Excitons, trions, biexcitons, and exciton-trion complexes
%many-body exciton complexes 
in two-dimensional transition metal dichalcogenide sheets
of MoS$_2$, MoSe$_2$, MoTe$_2$, WS$_2$ and WSe$_2$
are studied by means of
density functional theory and path integral Monte Carlo method in order to 
accurately account for the particle-particle correlations.
%using an approach
%that combines density functional theory and path integral Monte Carlo
%method, which exactly accounts for the particle-particle correlation.
In addition, the effect of dielectric environment on the
properties of these exciton complexes is studied by modifying the
effective interaction potential between particles.
Calculated exciton and trion binding energies are consistent 
with previous experimental and computational studies, and
larger systems such as biexciton and exciton-trion complex
are found highly stable.
Binding energies of biexcitons are similar or higher than
those of trions, but the binding energy of the trion depends
significantly stronger on the dielectric environment than that of
biexciton. Therefore, as a function of an increasing dielectric
constant of the environment the exciton-trion complex ''dissociates'' to a
biexciton rather than to an exciton and a trion.
\end{abstract}

\maketitle

%\section{Introduction}

Layered transition metal dichalcogenides (TMD)
are chemically, thermally, and mechanically stable even in the
monolayer form, and thus, provide an ideal platform for studying
condensed matter physics in two dimensions.
The semiconducting TMDs present many unusual optical properties
such as strong excitonic effects \cite{Mak10_PRL},
valley-dependent circular dichroism \cite{Mak12_NNano},
and second-harmonic generation \cite{Kumar13_PRB},
whose magnitude depends sensitively on the number of layers.
For instance, the prototypical MoS$_2$ material
is a semiconductor with 1.1 eV indirect band gap
in bulk, but 1.9 eV direct band gap in the monolayer \cite{Mak10_PRL}.
%This optical gap is related to excitonic transitions,
%with the fundamental gap being much larger.
Importantly, the reduced dimensionality is manifested
in a large exciton binding energy of 0.5--1 eV,
but also of significant binding energy in the case of
charged excitons, or trions, consisting of three charge carriers.
This suggests that even larger complexes might be stable.
Indeed, first experimental reports assigned to biexciton formation
have very recently appeared in the literature
\cite{Mai14_NL,Sie_arxiv,You15_NPhys}.
% but so far
%only a few isolated reports about observation of biexcitons and
%exciton-trion pair have been reported [Viitteet!!].
%but not very solid evidence...

Theoretical studies are invaluable in predicting the stability of
these complexes and in interpreting the experimental results.
Excitons can be calculated reliably from first-principles by solving
the Bethe-Salpeter equation (BSE) on top of quasi-particle band
structure. Binding energies have also been
calculated using simple variational or tight-binding
models based on an effective 2D
interaction potential and the effective mass approach
\cite{Keldysh79,Cudazzo11_PRB,Berkelbach13_PRB,Berghauser14_PRB,Zhang14_PRB,Chernikov14_PRL,Ganchev15_PRL,You15_NPhys},
yielding fairly good agreement with experiments and with the BSE
results in the case of excitons.
This has also raised interest to apply similar approaches to study 
larger exciton complexes \cite{You15_NPhys,QMCarxiv}, in comparison
to the theoretical estimates based on quantum well systems 
\cite{Kleinman83_PRB,Thilagam14_JAP,Mai14_NL}.

Difficulties in constructing reasonable wave function ansatz in the
case of the larger complexes hinders straightforward extension of
the simple variational models. Within the effective mass approach,
quantum Monte Carlo (QMC) methods, such as diffusion Monte Carlo and
path integral Monte Carlo (PIMC), provide accurate and powerful means
for studying few-particle systems
\cite{PhysRevA.86.052506,PhysRevA.90.042507,PhysRevA.91.062503}.  The
main advantage in QMC methods is the exact account of
particle-particle correlations, which is particularly important in
accurate description of exciton complexes. To this end, we utilize the
PIMC method, which is a basis set free approach for solving finite
temperature quantum statistics.

Although 2D materials are not directly bonded with the environment,
due to their thinness, they are highly sensitive to
electromagnetic fields, doping, or dielectric screening of their
surroundings. In particular, the Coulomb interaction between an electron
and a hole in exciton is screened by the dielectric environment and the
binding energy changes dramatically
\cite{Wirtz06_PRL,Komsa12_PRB2,Lin14_NL,Chernikov14_PRL,Zhang14_PRB,Berghauser14_PRB}.
It is rather surprising
then that the effect of dielectric environment on the binding energy
of trions, let alone on the larger complexes, has
been rarely investigated \cite{QMCarxiv}.
%not been investigated.

Here, we present the results from PIMC simulations for exciton, trion,
biexciton, and exciton-trion complexes for a set of the most common
layered TMD materials: MoS$_2$, MoSe$_2$, MoTe$_2$, WS$_2$, and WSe$_2$. We
focus on the binding energies and mean particle distances.  Our
approach is based on the effective 2D interaction and effective
masses, for which relevant parameters are calculated using density
functional theory (DFT). In addition, the effect of environment is
accounted for in the interaction potential, which allows us to
demonstrate the effect of the surroundings to the binding energies and
mean distances.
%We find that the biexciton is highly stable with binding energies
%close to or higher than those of trions.
%On the other hand, the binding energies of trion depends
%much more strongly on the dielectric environment.

%\section{Methods}

\begin{figure}[!ht]
\begin{center}
  \includegraphics[width=8cm]{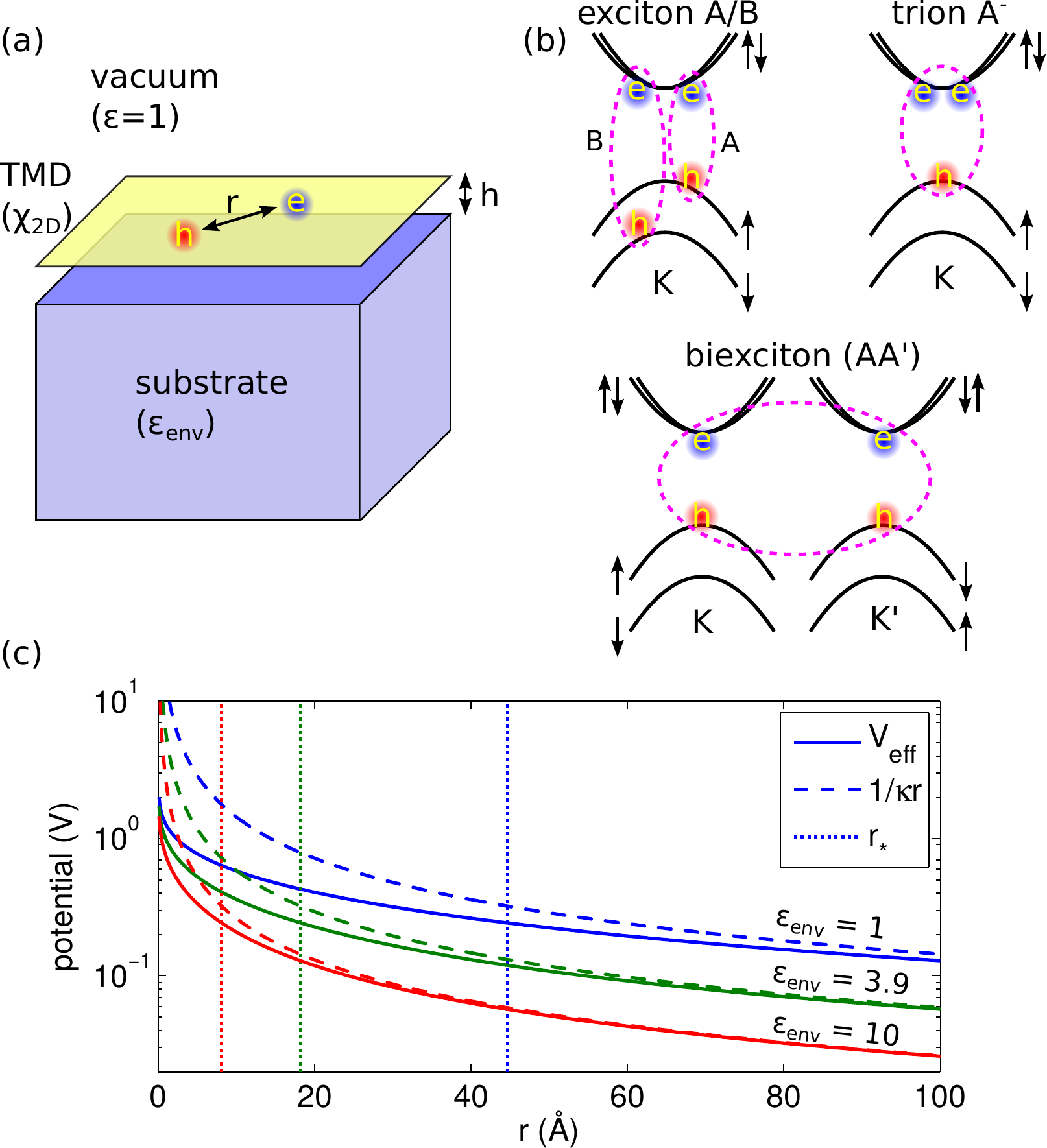}
\end{center}
\caption{\label{fig:system}
(a) Schematic illustration of the system geometry.
(b) Illustration of the exciton, trion, and biexciton systems
constructed from the electrons on the conduction band and holes
in the valence band.
(c) Effective interaction potential in the case
of $\epsenv{}=1$, $3.9$ (silica), and $10$ (sapphire), as compared to 
Coulomb interaction in homogeneous system scaled by $\kappa$.
}
\end{figure}

The geometry of the system is schematically shown in Fig.\ \ref{fig:system}.
The electrons and holes are confined to the 2D sheet placed on top
of a substrate.
We consider two-dimensional many-body Hamiltonians where the interaction 
potential between charged particles is given as in
Refs.~\onlinecite{Berkelbach13_PRB,Rodin14_PRB,Berghauser14_PRB}
%, that is, we have 
\begin{align}
\label{Eq:V}
V(r_{ij}) =\frac{q_iq_j\pi}{2\kappa r_{*}}\left[ 
H_0\left(\frac{r_{ij}}{r_{*}}\right)-Y_0\left(\frac{r_{ij}}{r_{*}}\right)\right],
\end{align}
where $H_0$ and $Y_0$ are the Struve function and the Bessel function of 
the second kind, respectively, $q_i$ represents the charge of the particle 
(here $q_i=e$ or $q_i=-e$).
The length scale in the presence of dielectric environment 
is given as
\begin{equation} \label{Eq:r*}
r_* = 2\pi \chi_{\rm 2D}/\kappa .
\end{equation}
$\chi_{\rm 2D}$ is the 2D polarizability of the sheet
\cite{Cudazzo11_PRB,Berkelbach13_PRB},
which can be evaluated from the bulk in-plane dielectric constant 
$\varepsilon_{||}$ and the layer separation $d_s$
(here half of the perpendicular lattice constant
since all considered materials have two layers within
the primitive cell):
$\chi_{\rm 2D} = d_s (\varepsilon_{||}-1)/4\pi$.
$\kappa$ is the average dielectric constant of the environment.
Here we consider the experimentally most relevant case of TMD
sheet placed on a substrate of dielectric constant $\epsenv$
and vacuum or air on the other side, in which case
$\kappa = (1+\epsenv)/2$.
%Distance of the sheet from the substrate $h\to 0$.
%The interaction potential is shown in Fig.\ \ref{fig:system}(c)
%for three representative values of $\kappa$.
The interaction potentials for three representative values of $\kappa$
in Fig.\ \ref{fig:system}(c) show strong screening by the 2D sheet
at short distances and approaching $1/\kappa r$ at $r \gtrsim r_*$.
We note, that this type of potential is only valid when 
$\kappa < \varepsilon_{||}$ \cite{supplement}.

The four considered exciton complexes are illustrated in
Fig.\ \ref{fig:system}(b).  In monolayer TMDs, due to the spin-orbit
coupling and the lack of inversion symmetry, the valence band maximum
(VBM) at K-point is split to two bands with distinct spin orientation.
The spins are further coupled to the valley index (K or K').  Thus, if
the many-body system is supposed to have two holes of opposite spins
in the topmost valence band state, they must be located in different
valleys. The conduction band minimum (CBM) is nearly degenerate (here
assumed degenerate) and there is no such restrictions for the
electrons.
% $\uparrow$ and $\spinarrow$

For the model interaction potential two material specific
parameters are then needed: polarizability of the sheet and
the effective masses of electrons and holes.
Both of these quantities can be obtained quite reliably from
DFT calculations. Our calculated numbers are collected in
Table \ref{tab:data}.
The atomic structures are optimized using revB86b-DF2 functional
\cite{Hamada14_PRB}, which yields structural parameters 
in very good agreement with the experiment \cite{Bjorkman14_JCP}.
Using these structures, the effective masses and dielectric constants
are then calculated using the PBE functional.
Bulk dielectric constants are calculated using density functional 
perturbation theory.
Since all materials considered here have direct gap 
with VBM and CBM located at K-point (in the monolayer form),
the effective masses are obtained from the monolayer structure
by fitting parabolas to the K-valley
and accounting for the spin-orbit coupling \cite{supplement}.

% Our effective masses for monolayer MoS2 are in good agreement with
% those given Habenicht15_PRB obtained from A and B exciton dispersion.

\begin{table}[!ht]
%\begin{center}
\caption{\label{tab:data} Material parameters needed for the effective
  mass theory description of the exciton complexes. 
  Layer separation $d_s$, and $\chi_{\rm 2D}$ are in {\AA}.
  The two values for $m_h$ correspond to the two spin-orbit split valence bands
  at K (first for the highest band). $\varepsilon$ is dielectric constant
  in directions parallel to the plane and normal to the plane.
}
\begin{ruledtabular}
%\begin{tabular}{lccccccc}
%\begin{tabular}{\textwidth}{@{\extracolsep{\fill}} lccccc}
\begin{tabular}{@{\extracolsep{\fill}} lccccc}
%\hline\hline
%Material  & $a$ & $d_s$ & $m_h$ & $m_e$ & $m_e/m_h$ & $\varepsilon$ & $\chi_{\rm 2D}$ \\
Material  & $d_s$ & $m_h$ & $m_e$ & $\varepsilon$ & $\chi_{\rm 2D}$ \\
\hline
MoS$_2$  & 6.180 & -0.54/-0.61 & 0.47 & 15.46/6.46 & 7.112 \\
MoSe$_2$ & 6.527 & -0.59/-0.69 & 0.55 & 17.29/7.95 & 8.461 \\
MoTe$_2$ & 7.054 & -0.62/-0.75 & 0.57 & 21.87/11.02 & 11.715 \\
WS$_2$   & 6.219 & -0.35/-0.49 & 0.32 & 13.92/5.98 & 6.393 \\
WSe$_2$  & 6.575 & -0.36/-0.53 & 0.34 & 15.47/7.22 & 7.571 \\
%\hline\hline
\end{tabular}
\end{ruledtabular}
%\end{center}
\end{table}

The path integral Monte Carlo simulations are carried out at $T=10$ K
using the effective mass approach with the masses obtained from our
DFT calculations, and the effective interaction potential given in
Eq.~\eqref{Eq:V}.  Apart from the exciton-trion complex we can
consider our particles as "boltzmannons", i.e., they obey the
Boltzmann statistics and are treated as distinguishable particles. In
case of negatively charged trions this is possible by assigning
spin-up to one electron and spin-down to the other one. With
biexcitons we simply apply the same for the positive particles, also.
The exciton-trion complex requires the account of Fermi statistics,
which in this work is dealt with by the restricted path integral Monte
Carlo approach, and the free particle nodal restriction
\cite{FermionNodes}.

In this work the statistical standard error of the mean with $2\sigma$
limits is used as an error estimate for all observables from our PIMC
simulations. Sampling in the configuration space is carried out using
the Metropolis procedure \cite{Metro53} with multilevel bisection
moves \cite{Chakravarty98}, and the thermal estimator
\cite{Ceperley95} is used in the calculation of the total energy.
We employ the PIMC method with the primitive approximation
\cite{Ceperley95}, for which we find that $T = 10$ K describes the
ground states of our systems accurately, and that using Trotter number
$M = 4000$ yields good balance between feasible amount of computer
time and accuracy. As the Trotter number $M$ tends to infinity, the
exact many-body results are obtained, but high-accuracy is often found
with reasonable values of $M$.
More details are given in the
Supplemental Material (SM) \cite{supplement}.

%\section{Results}

\begin{figure}[!ht]
\begin{center}
  \includegraphics[width=8cm]{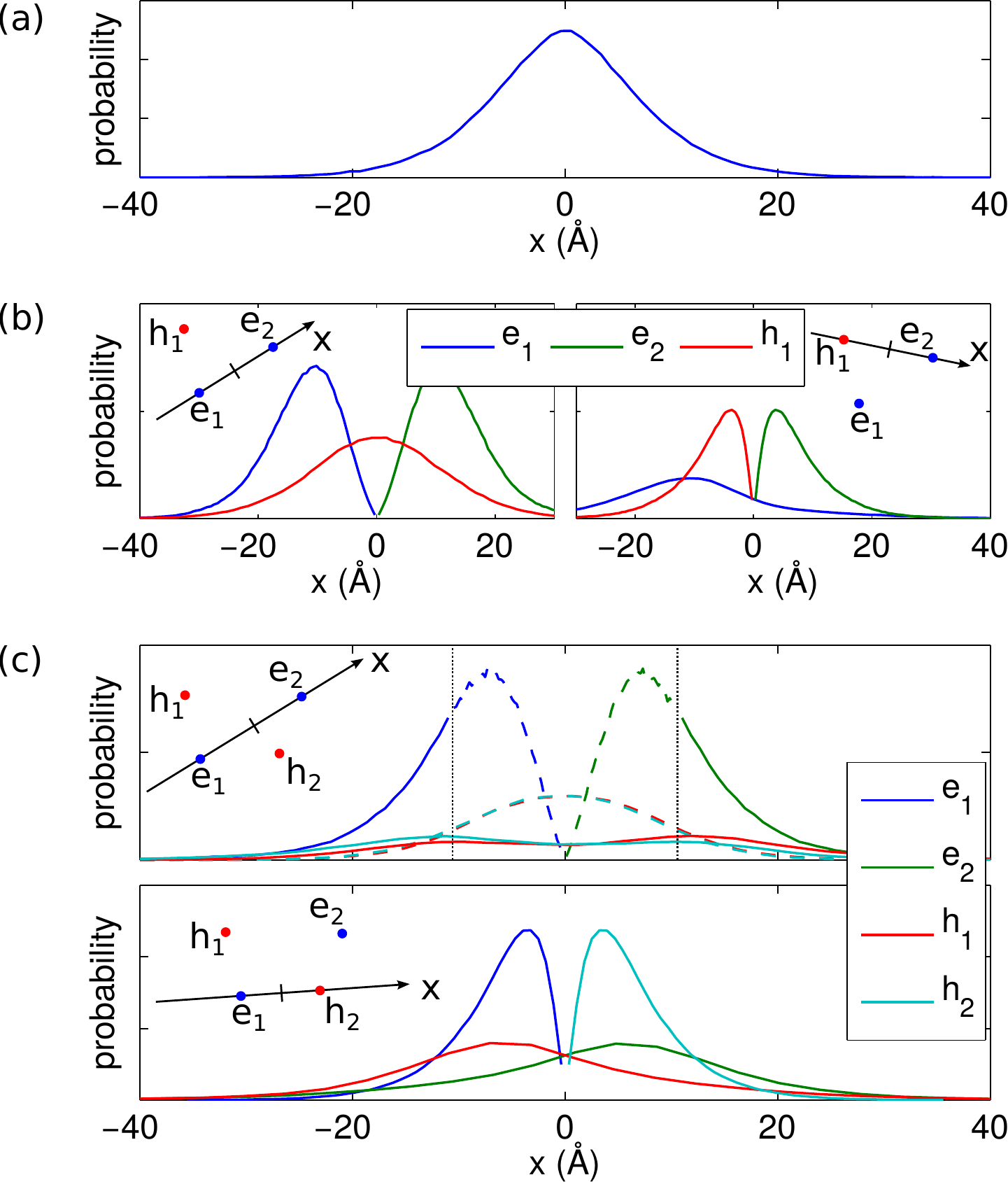}
\end{center}
\caption{\label{fig:morko}
Particle coordinate distributions in the case of (a) exciton,
(b) trion, and (c) biexciton. In (a), the hole is fixed at the origin.
In (b) and (c), $x$-axis is chosen to pass along two particles as
illustrated. Origin is at the center-of-mass of these two particles.
In (c), the distributions for electron-electron distance
larger (solid lines) and smaller (dashed lines) than 40 bohr
are distinguished.
}
\end{figure}

Before discussing the binding energies, it is useful to illustrate the
spatial distribution of the electrons and holes in these systems.  In
Fig.\ \ref{fig:morko}, we show the $x$-coordinate distribution of all
particles, when $x$-axis is chosen to pass through two of the
particles with origin at their center-of-mass.
In exciton, the electron
distribution when hole is fixed at the origin is similar to that of
hydrogen $1s$ state, showing exponential decay at larger distance.  In
the case of negative trion, the hole is largely located between the
electrons.  When the distance between electrons in biexciton is large,
there is one hole located close to each electron (solid lines in
Fig.\ \ref{fig:morko}c). 
%(The situation is identical when holes are chosen as fixed references.)
When the distance between electrons is small (dashed lines in 
Fig.\ \ref{fig:morko}c), the extent of the hole wave functions becomes
too large to make such distinction any more.

The root-mean square (rms) averaged electron-hole separation for exciton in 
suspended MoS$_2$ is found to be 11 {\AA},
%9.5 {\AA} (11.1 {\AA} rms distance),
which is in agreement with previous model calculations
\cite{Berkelbach13_PRB,Zhang14_PRB}
and with the $GW$+BSE calculated rms radius of 1 nm
reported in Ref.\ \onlinecite{Qiu13_PRL}.
%Such extent is at the limit of the applicability of the effective mass
%approach and can explain the underestimation of the exciton binding
%energy.
%The inter-particle separations for the 3--5 particle complexes are clearly
%larger, and the effective mass approximation should be safe.
The electron-electron rms separation for A$^-$ is 29 {\AA}
and for biexciton 23 {\AA}
(see SM for tabulated data of inter-particle distances in all
the considered TMDs \cite{supplement}).

The binding energies for all complexes are given in Table \ref{table:BE},
together with the experimental results.
Comparing the numbers over different materials, we observe that
the binding energy depends strongly on the material polarizability $\chi_{\rm 2D}$,
but weakly on the effective masses. As a consequence,
the results for A and B cases are generally very similar.
For instance, even if the effective mass ratio in WSe$_2$ differs
by more than 30\%, the binding energy differs less than 5\%.
Our calculated results for $B^-$ and $B^+$ trions and BB biexcitons
were also very similar to the $A$ counterpart and thus omitted from
Table \ref{table:BE}.
Finally, the effective mass insensitivity also extends to negative
and positive trions having nearly identical binding energies.

When comparing to the results obtained using ab initio $GW$+BSE approach,
as collected to Table \ref{table:BE}, our model tends to underestimate 
the exciton binding energies by $0.1$--$0.2$ eV; a satisfactory agreement.
The difference can arise from the fairly
small extent of the exciton wave function, and consequently breakdown
of the effective mass approximation. There are no first principles
results available for comparison in the case of trion, biexciton, or
exciton-trion complex. Nevertheless, our results for excitons and trions
are close to those obtained using the variational model 
for the case of $\kappa=1$ \cite{Berkelbach13_PRB}, but clearly smaller
than the estimate for biexciton given in Ref.\ \onlinecite{You15_NPhys}.
%thereby verifying their trial wave function.

%The ratio between biexciton and exciton binding energies are about 0.05
%close to that found for 3D, rather than 2D (in quantum wells) \cite{Kleinman83_PRB}.

\begin{table*}[!ht]
%\setlength{\extrarowheight}{1pt}
%\begin{threeparttable}
\caption{\label{table:BE} Binding energies for all considered systems both
in vacuum and in dielectric environment described by $\kappa=2$.
Experimental and computational results from literature are collected
for comparison.
Exciton energies are given in eV and binding energies of exciton complexes
are given in meV. The $2\sigma$ statistical error estimate is given in
parentheses for the PIMC results.}
\begin{tabular*}{\textwidth}{@{\extracolsep{\fill}} lccccc}
\hline\hline
  & MoS$_2$ & MoSe$_2$ & MoTe$_2$ & WS$_2$ & WSe$_2$ \\
\hline
 Exciton A & $0.5265(2)$ & $0.4769(2)$ & $0.3752(2)$ & $0.5098(2)$ & $0.4564(2)$ \\
 Exciton B & $0.5339(2)$ & $0.4853(2)$ & $0.3828(2)$ & $0.5309(2)$ & $0.4777(2)$ \\
 Other calc. & 0.55 \cite{Huser13_PRB}, 0.7 \cite{Lin14_NL}, 0.86 \cite{Berghauser14_PRB}
                            & 0.65 \cite{Ugeda14_NMat} & & 0.7 \cite{Ye14_Nat}   & \\
 Exciton A at $\kappa=2$ & $0.3486(2)$ & $0.3229(2)$ & $0.2608(2)$ & $0.3229(2)$ & $0.2946(2)$ \\
 Expt./Other calc.     & 0.43 \cite{CZhang14_NL,Hill15_NL}, 0.46 \cite{Berghauser14_PRB} & 
                                 & 0.58 \cite{Yang15_ACSNano} & 0.32 \cite{Hill15_NL} & 0.37 \cite{He14_PRL} \\
\hline
 Trion A$^-$ & $32.0(3)$ & $27.7(3)$ & $21.0(2)$ & $33.1(3)$ & $28.5(3)$ \\
 Trion A$^+$ & $31.6(3)$ & $27.8(3)$ & $20.9(3)$ & $33.5(4)$ & $28.5(4)$ \\
 Trion A$^-$ at $\kappa=2$ & $24.7(3)$ & $22.1(3)$ & $17.1(2)$ & $24.3(3)$ & $21.5(3)$ \\
 Expt.        & 18 \cite{Mak13_NMat} & 29 \cite{Ross13_NComm,Singh14_PRL,Liu15_2DM}
                       & 27 \cite{Yang15_ACSNano} & 34 \cite{Zhu15_SRep} & 31 \cite{Jones13_NNano,Liu15_2DM} \\
% Trion $B^-$ & $32.3(3)$ & $28.2(3)$ & $21.3(2)$ & $34.2(3)$ & $29.7(3)$ \\
% Trion $B^+$ & $32.3(3)$ & $28.2(3)$ & $21.1(3)$ & $33.9(4)$ & $29.5(4)$ \\
% expt        & 
%
\hline
 Biexciton AA & $22.7(5)$ & $19.3(5)$ & $14.4(4)$ & $23.9(5)$ & $20.7(5)$ \\
 Biexciton AA at $\kappa=2$ & $20.3(5)$ & $17.4(4)$ & $12.9(4)$ & $20.9(5)$ & $18.1(4)$ \\
% Biexciton $BB$ & $23.0(5)$ & $19.7(5)$ & $14.5(4)$ & $25.2(5)$ & $21.4(5)$ \\
% expt & & & & & \\
%
\hline
 Ex+Trion AA$^-$ & $17.0(6)$ & $16.4(5)$ & $12.5(5)$ & $14.9(6)$ & $14.9(6)$ \\
 Ex+Trion AA$^-$ at $\kappa=2$ & $13.5(4)$ & $12.7(4)$ & $10.0(4)$ & $13.1(5)$ & $12.2(4)$ \\
% Ex+Trion & $16.2(6)$ & $15.3(6)$ & $11.5(5)$ & $13.4(6)$ & $12.5(6)$ \\
% expt & & $4\pm 1.5$ singh14 & & & \\
\hline\hline
\end{tabular*}
\end{table*}

%2) sidosenergiat kappan (eps_env:n) suhteen

\begin{figure}[!ht]
\begin{center}
  \includegraphics[width=8cm]{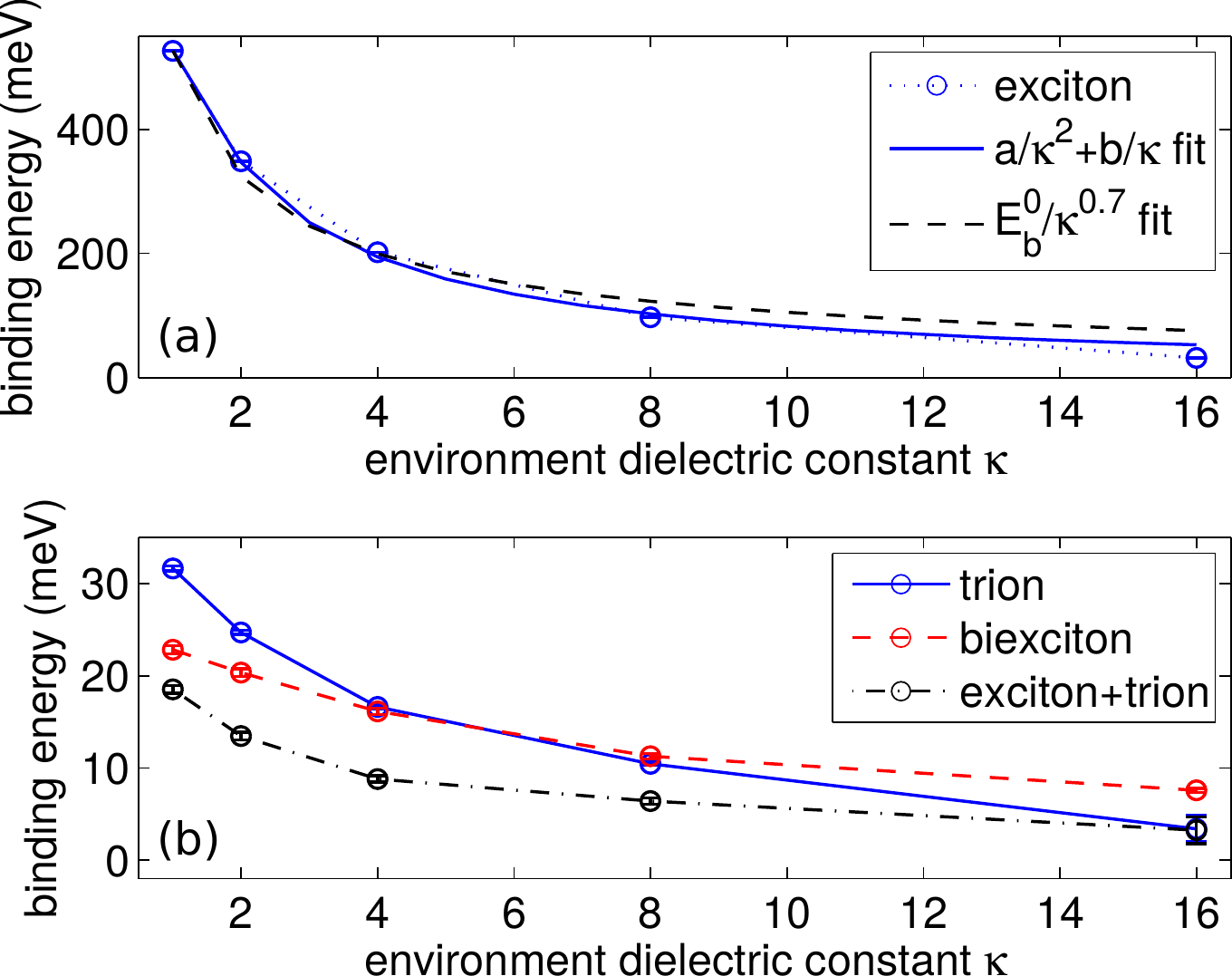}
\end{center}
\caption{\label{fig:kappa}
(a) Calculated exciton binding energies as a function of $\kappa$
together with error bars ($2\sigma$). Dotted line is guide for the eye.
Two fits are also shown: The coefficients for the first
(solid line) are $a=-325$ meV and $b=853$ meV.
%       a =        -325  (-497.8, -152.3)
%       b =       852.8  (698.2, 1007)
The second fit (dashed line) was proposed in Ref.\ \onlinecite{Lin14_NL}.
(b) Binding energies and error bars for trion, biexciton,
and exciton-trion complex as a function of $\kappa$.
}
\end{figure}

In experiments, TMD sheets are rarely suspended in vacuum.
Dielectric environment has strong effect on the screening of the
interactions within the sheet and consequently on the binding energies
\cite{Komsa12_PRB2,Lin14_NL,Chernikov14_PRL,Zhang14_PRB,Berghauser14_PRB}.
The binding energies as a function of the average dielectric constant
of the environment $\kappa$ are shown in Fig.\ \ref{fig:kappa}.
For excitons, the dependence on $\kappa$ can be fitted reasonably well
with $E_b(\kappa=1)/\kappa^\alpha$, when $\alpha\approx 0.7$
in agreement with the asymptotic form found in Ref.\ \onlinecite{Lin14_NL}.
With increasing $\kappa$, the trion binding energy decreases faster 
than that of biexciton and they are found equal at $\kappa\approx 4$.
Since the short range interaction is more strongly affected
by the dielectric environment (cf.\ Fig.\ \ref{fig:system}c),
and considering the trion geometry shown in Fig.\ \ref{fig:morko}b,
the repulsive electron-electron interaction is affected less
than the attractive electron-hole interaction.
If biexciton is approximated as two weakly bound excitons, then their
binding energy should have only weak $\kappa$-dependence.

Comparing our calculations to experimental results is hindered by
the large variations in the reported numbers.
%due to different conditions, substrates, experimental methods.
Fortunately, however, many of these experiments are carried 
out with the TMD sheet placed on a SiO$_2$ substrate. Then, in order 
to facilitate the comparison we have recalculated all binding energies
for $\kappa=2$, approximately corresponding to a situation of SiO$_2$ on
one side and vacuum/air on the other.
The results are given in Table \ref{table:BE}.
In the case of excitons, our calculated binding energies are now
within 0.1 eV of the experimental values.
For trions, our calculations are also generally in line with the
experimental results or slightly underestimated as in the case of excitons.
It is worth noting though, that some authors have reported similar binding 
energies for the A$^-$ and A$^+$ trions \cite{Ross13_NComm} % MoSe$_2$
in agreement with our results, 
whereas others have found larger binding energy for the A$^-$ than for the
A$^+$ \cite{Jones13_NNano,Ross14_NNano}.

In contrast to excitons and trions, which are commonly observed,
the reports for biexcitons are very scarce.
You \emph{et al}.~ report binding energy of 54 meV for both inter- and
intravalley biexcitons in WSe$_2$ \cite{You15_NPhys},
Mai \emph{et al}.~ report binding energy of 70 meV for 
AA' biexciton (where A' denotes exciton in the K' valley) in MoS$_2$
% on Al$_2$O$_3$ (again $\kappa\approx 2$)
%and relatively long decay time of 10 ps 
\cite{Mai14_NL}, and
%See SM for analytic estimates (depending on the dielectric constant
%of the environment): exciton 0.25--1 eV, biexciton 35--140 meV
Sie \emph{et al}.~ found binding energy of 40 meV for AB biexciton, 60 meV for AA' biexciton
in MoS$_2$ \cite{Sie_arxiv,privcomm}.
%[Large binding energy difference indicates that exchange may be important.]
These experimental results are clearly larger than our calculated
values of about 20 meV.

Likely explanation to the discrepancy is the neglect of exchange,
and especially the electron-hole exchange, in our
calculations. $GW$+BSE calculations yield dark exciton 20 meV
% (but 34 meV for B and 25 meV if no SOC) 
below the bright one
%(extracted from the calculation reported in Ref.\ \onlinecite{Komsa13_PRB2}).
\cite{Qiu_arXiv,Komsa13_PRB2},
which originates from electron-hole exchange due to
vanishing splitting of the conduction band in MoS$_2$.
%that is then missing from the energy of our 
%For exciton, there is no contribution, since
%only the opposite spin-case ($h_\downarrow e_\uparrow$) is optically bright.
This is the exchange energy that should be added to our calculated value
to yield the bright exciton energy.
% 20 meV given for MoS2 in Qiu arXiv 1507.03336
% Similar values were calculated in Forney74_NC 
Neglect of exchange between electrons in negative trion, i.e., setting
their spins the opposite, inevitably leads to missing the exchange
from one electron-hole pair, with energy contribution similar to
an exciton or smaller since the electron and hole are more separated.
%, the electron-hole exchange contribution is expected
%to be smaller than for exciton.
%
In AA' biexciton there is no exchange between electrons or between holes,
since they are assumed to have opposite spins.
The electron-hole pairs at K or K' valley should have similar 
electron-hole exchange contributions as excitons.
However, if we consider that the electron at K is bound to hole at K'
and vice versa, these are dark excitons with no electron-hole exchange.
Comparing such configuration to the energies of two bright excitons
that are missing exchange energies in our calculations, leads to 
total energy correction of 40 meV.
%However, starting from excitation of spin-$\uparrow$ electron
%($h_\downarrow e_\uparrow$ A exciton), and when exciting additionally
%spin-$\downarrow$ electron (A' exciton, also optically allowed),
%the biexciton may reorganize to assume a configuration consisting of
%$h_\downarrow e_\downarrow$ and $h_\uparrow e_\uparrow$ excitons,
%i.e., two dark excitons. Such configuration should experience
%almost twice the electron-hole exchange of exciton, about 26 meV. 
Thus obtained binding energy of 60 meV is then in line with the
experimental values.
Another explanation could be that in the experiment the biexcitons are
bound to e.g.\ impurities.

%Since the diffence in the binding energies of trion and biexciton,
%as seen in Fig.\ \ref{fig:kappa}, is less than 5 meV for nearly all values 
%of $\kappa$, they are likely difficult to distinguish in some experiments.

The only report of exciton-trion complex, to the best of our
knowledge, is in Ref.\ \onlinecite{Singh14_PRL}. They deduced binding
energy of $4\pm 1.5$ meV for MoSe$_2$ on SiO$_2$. Our number from
Table \ref{table:BE} for this case is $12.7(4)$ meV, which is
substantially larger.
% Interestingly, considering such an exciton-trion
%complex in which the holes have like spins (i.e., AB'$^{-})$, the
%binding energy from our PIMC simulations is roughly $1.8$ meV [Need to
%  continue the run for longer... let's see what comes out!!!], which
%is in reasonable agreement with the experimental value. Moreover, it
%also follows the trend of underestimating the binding energy.

\begin{figure}[!ht]
\begin{center}
  \includegraphics[width=8cm]{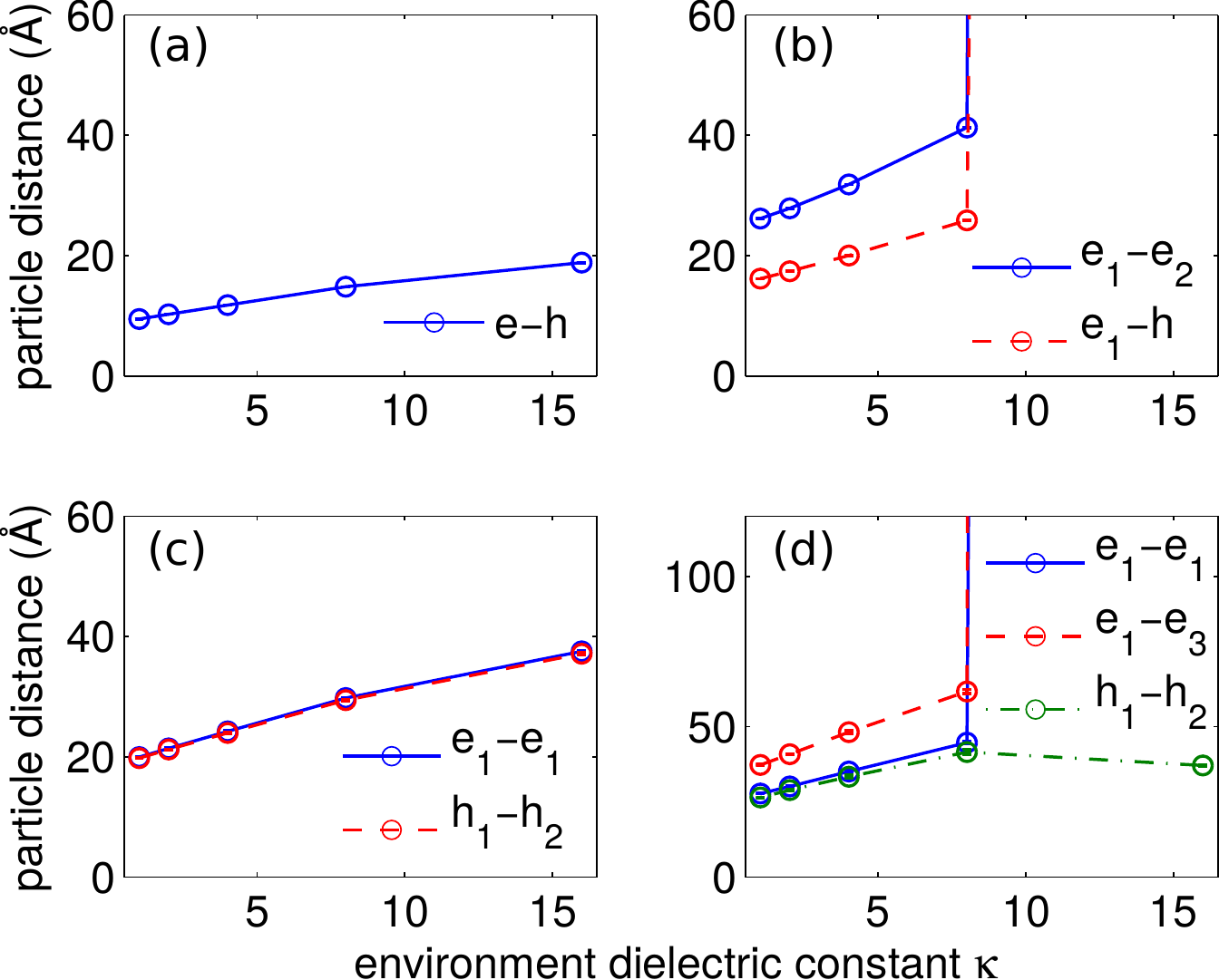}
\end{center}
\caption{\label{fig:dist}
Selected average inter-particle distances for exciton (a),
trion (b), biexciton (c), and exciton-trion complex (d)
in MoS$_2$ as a function of $\kappa$.
}
\end{figure}

Finally, to illustrate the dependence of the system size on the environment,
the inter-particle distances as a function of $\kappa$ are shown in
Fig.\ \ref{fig:dist}.  The particle distances generally scale linearly
with $\kappa$. 
% Exciton and biexciton are found to dissociate after
%$\kappa>16$, 
Trion and exciton-trion complex are found to dissociate after
$\kappa>8$. We note, that the dissociation is facilitated by the
non-zero temperature (10 K) used in our PIMC calculations.
Interestingly, the exciton-trion complex does not
dissociate to exciton and trion, but to biexciton (compare to
$h_1-h_2$ distance in biexciton) and a free electron. This is further
confirmed by inspection of other particle-particle distances, i.e.,
electron-hole distances (not shown in the figure).

%\section{Conclusions}

In conclusion, we have studied excitons, trions, biexcitons,
and exciton-trion complexes in two-dimensional transition metal
dichalcogenides by an approach combining density functional 
theory with quantum Monte Carlo method. We focused on the binding
energies, inter-particle separations, and on the role of dielectric
environment. Our approach
reproduced exciton and trion properties in reasonably good
agreement with experiment. We found that the larger complexes
should also be stable with binding energies comparable to those
of trions, although the relative stability can be controlled
by the dielectric environment of the 2D sheet.
Due to the large binding energies, environmental control, and
coupling with the valley and spin indices of the material, 
we expect TMD materials to provide a versatile ``laboratory''
for studying, experimentally and theoretically, the physics of 
correlated many-body systems going even beyond the 3--5 particle
complexes considered here. 

%\section{Acknowledgments}

We thank the Academy of Finland for the support under Project
No.~126205 (IK) and No.~263416 (HPK), and through its Centres of 
Excellence Programme (2012-2017) under Project No.~251748 (HPK). 
We also thank CSC--IT Center for Science Ltd.~ 
and Tampere Center for Scientific Computing for 
generous grants of computer time.

%\bibliography{refs}

\end{document}